\documentclass[useAMS,usenatbib]{mnras}

\usepackage{txfonts}
\usepackage{times}
\usepackage{latexsym, amssymb, amscd, psfrag}
\usepackage[dvipdfmx]{graphicx} 
\usepackage{bmpsize}
\usepackage{epsfig,color}
\usepackage{subfig}
\usepackage{morefloats,rotating,float}
\usepackage{gensymb}
\usepackage{multirow,array}
\usepackage{xcolor}
\usepackage{soul}

\usepackage[T1]{fontenc}
\usepackage{ae,aecompl}
\usepackage{multicol}

\usepackage{graphicx}    
\usepackage{amssymb}    
\graphicspath{{Figures/}} 
\usepackage{subfig}

\newcommand{\e}{{\sc eagle}}
\newcommand*{\Comb}[2]{{}^{#1}C_{#2}}%
\newcommand{\smass}{{$M_{*}$}}

\title[Galaxies on filaments]{Study of galaxies on large-scale filaments in simulations}

\author[Singh et al.]{
Ankit Singh,$^{1}$\thanks{E-mail: ansingh16@gmail.com}
Smriti Mahajan,$^{1}$
Jasjeet Singh Bagla$^{1}$
\\
$^{1}$Indian Institute of Science Education and Research, Mohali, Punjab, India\\
}

\date{Accepted XXX. Received YYY; in original form ZZZ}

\pubyear{}

\begin{document}
    
\captionsetup[subfigure]{labelformat=empty}
\renewcommand\thesubfigure{(\alph{subfigure})}

\label{firstpage}
\pagerange{\pageref{firstpage}--\pageref{lastpage}}
\maketitle

\begin{abstract}
We use data from the Evolution and Assembly of GaLaxies in their Environment ({\sc eagle}) cosmological simulation to study properties of galaxies in the cosmic web. Galaxies become more redder and form stars at a lower rate relative to their counterparts further away from the cylindrical axis of the large-scale filaments. These trends are particularly strong for galaxies with $M_*/M_{\odot}\lesssim10^{10}$. We also find that at distances $<0.5$ Mpc from the spine of the filaments, the median gas and stellar mass fraction in filament galaxies rises sharply with decreasing distance from the spine of the filament. These results, together with matching trends in the SFR/$M_*$ and the $g-r$ colour of filament galaxies suggest that (i) the intrafilamentary gas condenses into the filament galaxies thus fuelling star formation in them, and (ii) increased number density of galaxies closer to the central axis of the filament enhances the rate of gravitational interactions among filament galaxies closer to the spine. 
\end{abstract}



\section{Introduction}

 Galaxies form and evolve on the cosmic web: a complex network of intricately woven filaments and sheets of matter which intersect at the nodes, aka the clusters of galaxies \citep{1996bond}. The distribution of all matter on such a web-like structure is expected from the theory of structure formation \citep{zeldovich82} even before it was observed. However,
 the impact of the cosmic web on the properties of galaxies through their interactions with the large-scale environment is yet to be understood fully. 

 The role of the large-scale environment in modulating the evolution of galaxies has been well established in the literature 
 \citep{1997Dressler,1988Binggeli,2004Kauffmann,2009Weinmann,2012Peng,2016Boselli}. 
 While interactions with the ambient hot gas in dense environments such as clusters and groups can lead to a loss of cold gas from galaxies through ram-pressure stripping \citep{1972Gunn}, 
 gravitational interactions between galaxies \citep{2006Kewley,2015Sobral,2015Stroe}, or harassment of low-mass satellite galaxies by a more massive counterpart \citep{1996Moore} can quench 
 formation of stars in intermediate and low-density environments. 

 Simulations have confirmed that even though there are more galaxies in clusters, a large fraction of all galaxies resides on large-scale filaments \citep[e.g. see table~2 of][]{libeskind18}. 
 So in order to understand the role played by the cosmic web in the evolution of galaxies, it becomes critical to study the large-scale filaments and their influence on the 
 properties of galaxies. The intra-filament medium (IFM) is expected to be at $10^5-10^7$ K, and over-dense by a factor of $\sim 20$ relative to the mean of the Universe \citep[e.g.][]{cen99,croft01}. 
 Hydrodynamical simulations show that the Bremsstrahlung brightness in the cosmic web reaches at most $10^{-16}$ erg s$^{-1}$ cm$^{-2}$ arcmin$^{-2}$ \citep{dolag06}. Hence, while we 
 overcome the technical challenges to observe the IFM directly, it is essential to study its impact on the galaxies residing therein by studying their properties.                                              

 There is plenty of evidence in the literature to suggest that the galaxies on filaments have higher star formation rate relative to their counterparts in clusters and voids 
 \citep{porter07,fadda08,porter08,mahajan12,2018Smriti,vulcani19}. Furthermore, galaxies on filaments are found to be more metal-rich \citep{2015Darvish}, as well as host more satellite galaxies 
 \citep{guo15}. Studies of galaxy pairs on filaments indicate a significant alignment between the pair axis and the axis of the filament on which they reside
 \citep{lambas88,donoso06,mesa18}. In particular, \citet{mesa18} found that the alignment for pairs of elliptical galaxies is stronger than the pairs comprising one or both spiral galaxies, especially 
 within $200~h^{-1}$ kpc of the spine of the filament.     
 
 Several recent studies suggest that the properties of galaxies on filaments change as a function of their distance from the cylindrical axis of the filament. Using optical spectroscopic and 
 photometric data from the Galaxy And Mass Assembly (GAMA) survey and the Horizon-AGN simulation suite, \citet{2018Kraljic} found that galaxies tend to become passive closer to the 
 filament axis. Multi-wavelength data from the GAMA survey were also used by \citet{alpaslan16} to show that galaxies ($M_{*} > 10^9; z < 0.09$) become less star forming and more massive 
 with decreasing distance from the filaments' axis.  
 \citet{2018Smriti} studied the Coma supercluster ($z\sim 0.023$) using similar optical data from the Sloan Digital Sky Survey (SDSS) and ultraviolet (UV) data from the Galaxy Evolution 
 Explorer ({\it Galex}) mission. One of their major results is that the colours of galaxies become redder and emission in the H$\alpha$ line decreases on approaching the spine 
 of the filament. Similar results have also been obtained for galaxies at higher redshift \citep{chen17,malavasi17,laigle18,luber19}.
 
 In the last few years, state-of-the-art simulations have been employed to understand the symbiotic relationship between the large-scale structure and its galaxy constituents. Large cosmological 
 simulations not just serve as a test bed for validating various theories of structure formation and evolution, but also serve as a laboratory to predict the properties of the yet to be observed 
 constituents of the Universe. For instance, \citet{2018Martizzi} used the IllustrisTNG simulations to show that the filaments contain more star-forming gas, as well as higher mass 
 fraction of such gas relative to clusters. In a study based on the Horizon MareNostrum simulation, \citet{2010Gay} explored the properties of galaxies at redshift of $z = 1.5$ (box size of 
 $50 h^{-1}$ Mpc), and found that the {\it G-K} colour of galaxies becomes redder on approaching the spine of the filaments.

 Even though the present day simulations are able to produce global trends in the properties of galaxies, they depend upon recipes to mimic effects of various sub-grid physical processes at play. 
 Since these parameters can not be found using first principles, it is critical to understand the variation in properties of galaxies as a continuous function of their environment in order to 
 validate different sub-grid recipes. The Evolution and Assembly of GaLaxies and their Environment \citep[{\sc eagle};][]{EAGLE_paper,2015Crain,2016McAlpine,2016Tray,2017Camila} is one such suite of simulations 
 which has been used to study various aspects of evolution of galaxies in recent years \citep{2015Tray,Furlong2015,2017Rossi}. A very useful aspect of the \e~simulations is that the 
 calibration of the feedback parameters is performed by broadly matching the galaxy stellar mass function (GSMF) at $z \sim 0$, maintaining reasonable galaxy sizes. Through this new 
 implementation, \e~can reasonably match and predict multi-wavelength observations at large scales, making them an ideal resource to study galaxies in different environments. 
 
 We describe our methodology in the next section, followed by the results of our analysis in Sec.~\ref{sec:Results}, which are discussed in the 
 context of the existing literature in Sec.~\ref{sec:Discussion}. We finally conclude in Sec.~\ref{sec:Conclusion}.

\section{The \e~Simulations}
\label{sec:EAGLE} 

 In this work we employ data produced by the \e~simulation \citep{EAGLE_paper,2016McAlpine}, which is a suite of hydrodynamical simulations run with modified version of the Gadget3 code. 
 The  simulation  uses  a  flat  $\Lambda$CDM  cosmology  with  parameters  taken  from  the Planck mission \citep{2014Planck} results: $\Omega_\Lambda$ = 0.693, $\Omega_m$ = 0.307, $\Omega_{b}$ = 0.04825, $\sigma_8$ = 0.8288, $\mathrm{n_s}$ = 0.9611,Y = 0.248 and $\mathrm{H_0}$ = 67.77 $\mathrm{km \ s^{-1} \ Mpc^{-1}}$(i.e. $h$ = 0.6777). Specifically, we use the reference 
 model RefL0100N1504 with co-moving box length of 100 $h^{-1}$ Mpc. The total number of particles in the RefL0100N1504 run is $2 \times 1504^3$, giving an initial gas particle mass ($ m_{\rm g}$) of 
 $1.81 \times 10^{6} M_{\odot}$, and a dark matter particle mass of $9.70 \times 10^{6} M_{\odot}$, respectively. By using state-of-the-art numerical techniques, \e~ differs from other simulations in its implementation of energy feedback from stars, i.e. it does not distinguish between core-collapse supernovae, stellar winds and radiation pressure. In this study we have used the snapshot at $z = 0.1$ to create mock observational data.  

\subsubsection{The physical parameters}

 Star formation in the \e~simulations is implemented using the Kennicutt-Schmidt relation \citep{1998Kennicutt} between the gas mass density and the star formation density \citep{2008Schaye}. 
 In particular, the star formation rate decreases with increasing metallicity above the density threshold given by \citet{2004Schaye} as: 
\begin{equation}
n_{\rm H}^* = 10^{-1} {\rm cm}^{-3} \big(\frac{Z}{0.002}\big)^{-0.64} 
\end{equation}
where, {\it Z} is the gas metallicity. The simple stellar populations represented by these stellar particles follow the Chabrier initial mass function (IMF). Furthermore, in order to prevent 
artificial fragmentation, a density dependent temperature floor ($T_{\rm eos} ( {\rho}_{\rm g} )$) is applied following the equation of state : $P_{\rm eos} \propto {\rho}_{\rm g}^{{\gamma}_{\rm eos}}$
 with ${\gamma}_{\rm eos} = 4/3$, where $P$ is the pressure, $\rho_{g}$ is the gas density and $\gamma$ is the ratio of the heat capacities. The temperature floor is normalised to $T_{\rm eos}=8000$ K at $n_{H}=0.1 \ \mathrm{cm^{-3}}$ to mimic gas in cold phase. When the gas density exceeds $n_{\rm H}^*$,
 and the temperature $\log_{10} (T/{K}) < \log_{10} (T_{\rm eos}/{K}) + 0.5$, the gas particles are assigned a star formation rate (SFR):  

\begin{equation}
\label{eq:sfr}
{\dot m}_* = m_{\rm g} \ A \ (1 {\rm M}_{\odot} {\rm pc}^{-2})^{-n} \ (\frac{\gamma}{G} 
f_{\rm g} P)^{(n-1)/2} ,
\end{equation} 
 where, $G$ is the gravitational constant, $\gamma = 5/3$ is ratio of specific heats and  $f_{\rm g}$ is the fraction of mass in the gas, and is kept at 1.0 for the RefL100N1504 model. 
 Parameters $n=1.4$ and $A= 1.515 \times 10^{-4} \ M_{\odot} \ {\rm yr}^{-1} \ {\rm kpc}^{-2}$ are estimated from the fit to observational data \citep{1998Kennicutt}.  

 The rates of radiative cooling and heating are computed on an element-to-element basis in the presence of \citet*{2001Haardt} ultraviolet or x-ray ionizing background, and the cosmic microwave 
 background. Eleven elements (H, He, C, N, O, Ne, Mg, Si, S, Ca and Fe) which are important for calculating cooling rates at $T>10^4K$ \citep{2009Wiersma} along with the total metallicity are 
 tracked in the simulations. 
 
 The simulations incorporate various sub-grid feedback mechanisms including stellar winds from asymptotic giant branch (AGB) stars, supernovae \citep[both core collapse and type IA;][]{2009Wiersma1} and active galactic nuclei \citep[AGN;][]{EAGLE_paper}. The prompt stellar feedback is implemented in a manner similar to \citet{2012Dalla}. In this `stochastic feedback' model, the fraction of energy ($f_{\rm th}$) released by the core collapse supernovae is given by:
\begin{equation}
\label{eq:fth}
f_{\rm th} = f_{\rm th,min} + 
\frac{f_{\rm th,max}-f_{\rm th,min}}{1 + \big( \frac{Z}{0.1 Z_{\odot}} \big) ^{n_Z} 
    \big( \frac{n_{\rm H, birth}}{n_{\rm H,0}} \big) ^{-n_n}}  , 
\end{equation}
where, $f_{\rm th,max}$ and $f_{\rm th,min}$ are the asymptotic values of $f_{\rm th}$ and $n_{\rm H, birth}$ is the density of the parent gas particle. In this model, $n_n$, $n_Z$, and $n_{\rm H,0}$ are free parameters. This fraction is injected into the neighbouring cells 30 Myr after the birth of the stellar particle. The thermal energy ejected by a star is distributed stochastically isotropically, resulting in temperature change of $\Delta T = 10^{7.5}$K.

When the mass of a halo reaches $10^{10}M_{\odot}$, a gas particle at the centre of the potential well is converted into a seed black hole with a mass of $10^5$ ${\it{h^{-1}}}M_{\odot}$ \citep{2005SringelDiMatteo}. The accretion of matter on the central black hole is based on a modified Bondi-Hoyle model \citep{EAGLE_paper}. A fraction of 0.015 of the rest mass energy 
 of the matter accreted on the central black hole is returned to the surrounding thermally as a jump of $\Delta T = 10^{8.5}$K. 

All the free parameters related to physical processes are calibrated to reproduce the GSMF and galaxy sizes at $z = 0.0$.

\begin{figure}
    \includegraphics[scale=0.55]{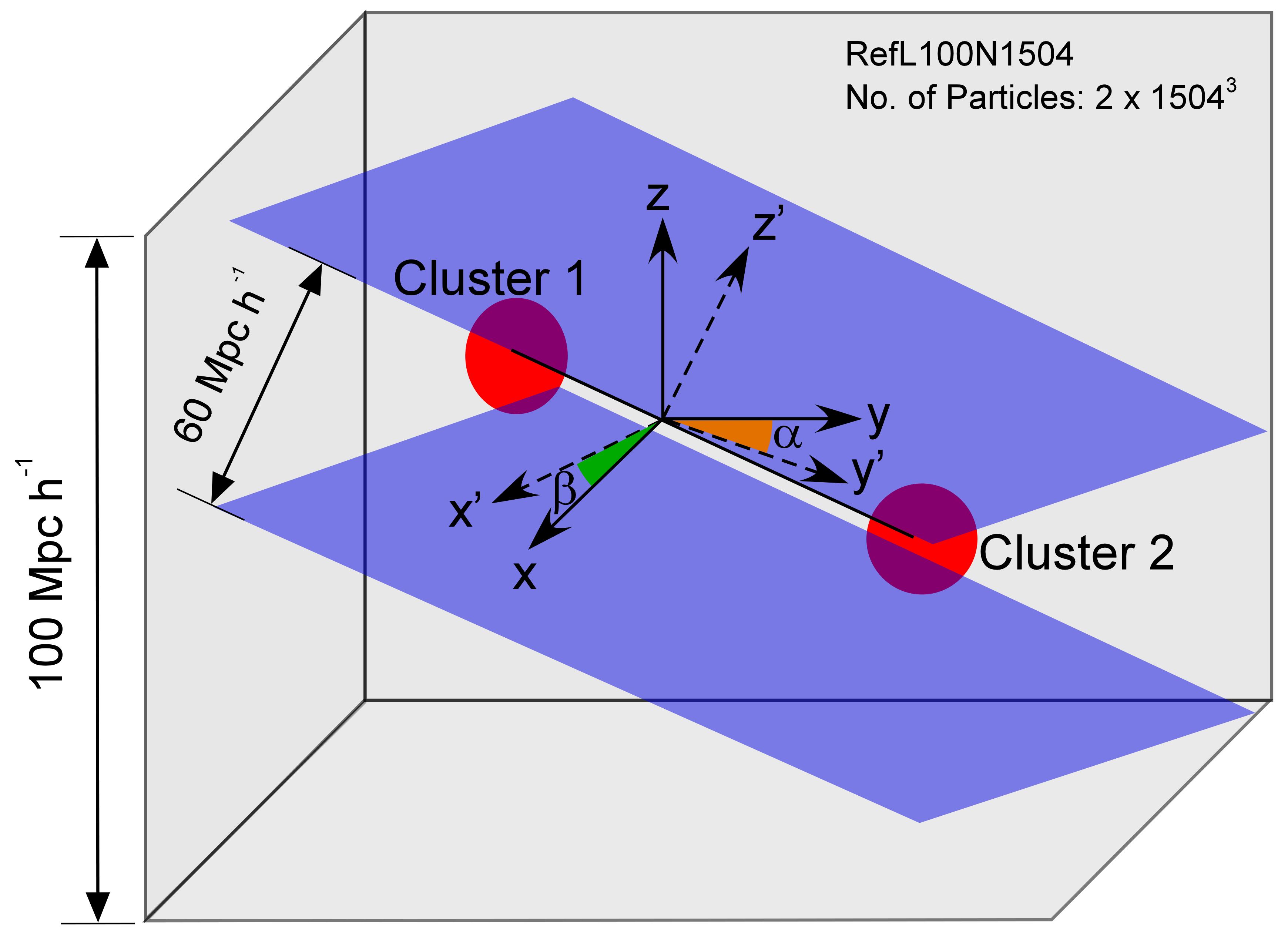}
    \caption{The setup employed for creating the slices for mock observations. A slice is cut from the simulation box around each pair of clusters ($M_{200}/M_{\odot} > 10^{14}$), by translating the original coordinate system ($O$), such that the centre of the modified coordinate system coincides with the centre of the line joining the two clusters. The system is then rotated about this axis, and angles $\alpha$ and $\beta$ are randomly generated. This new coordinate system is referred in the text as $O^{\prime}$. Throughout this paper, environment of galaxies is characterised based on the 2D projection of 36 such slices in the $O^{\prime}$ frame of reference.}
    \label{plot:setup}
\end{figure}

\begin{figure}
    \centering     
    {\includegraphics[scale=0.6]{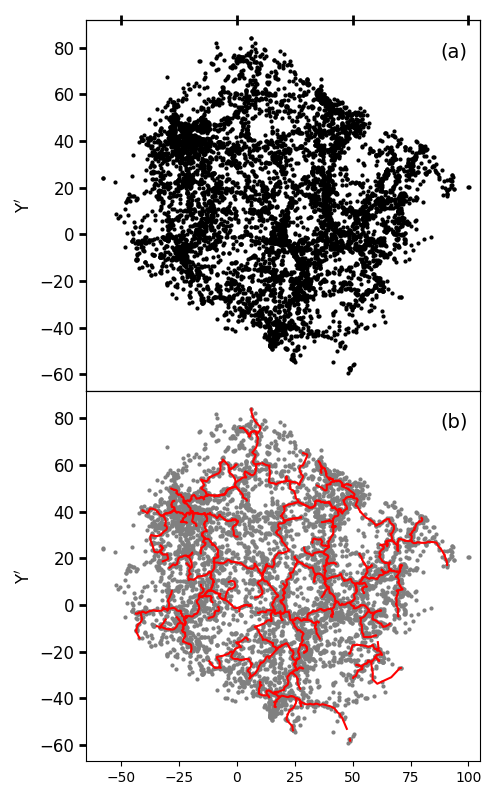}}
    \caption{(a)The projected position of galaxies in one of our slices.    
    (b) Same as (a), with {\it red lines} overplotted to  represent the filaments identified by {\sc disperse}.}
    \label{proplot}
\end{figure}

\begin{figure}
    \includegraphics[scale=0.4]{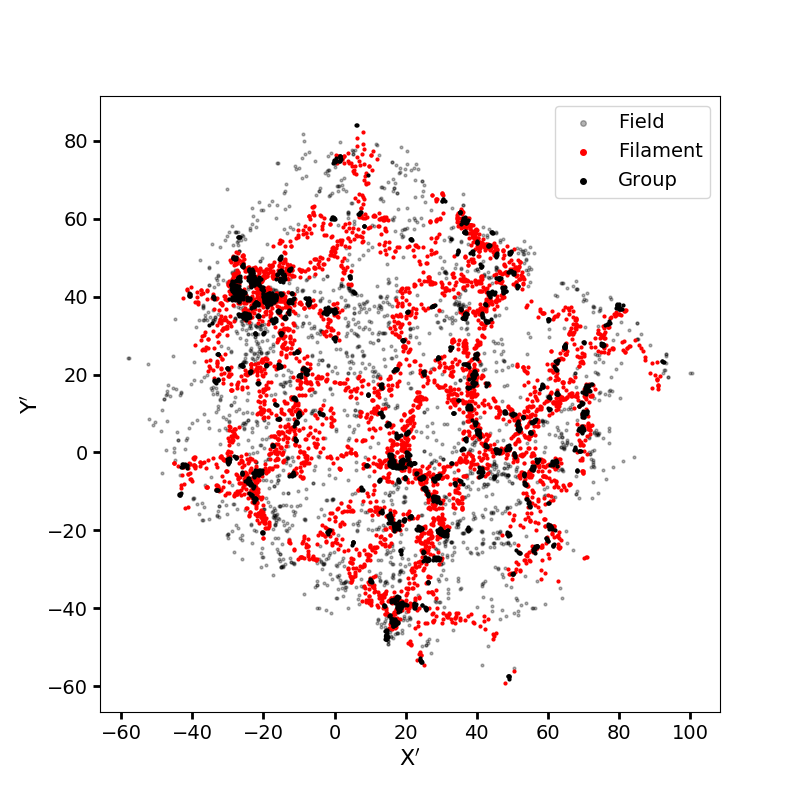}
    \caption{The projected position of galaxies in the same slice shown in Fig.~\ref{proplot}, with each point labeled with the environment it belongs to. The radius for the filaments is assumed to be 2 Mpc (see text for details). The {\it black, red}, and {\it grey points} represent galaxies in groups and clusters, filaments, and voids, respectively.}
    \label{finalexslice}
\end{figure}


\subsubsection{Properties of galaxies}
\label{Propsec}

In \e~simulations, the dark matter haloes are identified using the friends-of-friends (FOF) algorithm \citep{1984Einasto}, such that the dark matter particles separated by less than 0.2 times the mean 
inter-particle separation are bound into a single halo. The individual galaxies are identified using the SUBFIND algorithm \citep{2001Springel,2009Dolag}, which labels the galaxy 
 closest to the centre of the parent halo as the central galaxy, and other haloes inside the parent as satellite galaxies.

It is difficult to quantify the mass or luminosity of galaxies in simulations using the SUBFIND algorithm, because it assigns the mass of unbound particles to the central galaxy. This leads to an extended stellar distribution for such galaxies. Hence, in this work we use the stellar mass of the subhalo obtained within an aperture of 30 kpc in physical units (pkpc, henceforth), centred at the minima of the subhalo potential \citep{2015Tray,EAGLE_paper}. The choice of aperture is motivated by the fact that 3-D aperture of 30 kpc matches the results obtained using 2-D Petrosian aperture \citep{EAGLE_paper}. 

The \e~simulation uses the \cite{2003Bruzal} population synthesis models for each particle to obtain the spectral energy distribution (SED) of galaxies. The spectra convolved with the filter response function within 30 pkpc aperture gives the broadband colours of the galaxies in optical and near infrared wavebands \citep{2010Doi,2006Hew}. 
We employ some of these rest-frame colours to study the properties of galaxies. A full list of all the variables used in our analysis is provided in Table~\ref{proptable}. In this work, we use the SFR inside 30 kpc in physical units centred at minima of subhalo potential to explore such trends. The SFR has been measured using the mass of dense gas in the subhalo \citep{2008Schaye}.

 Studies of large-scale filaments based on observations mainly focus on the outskirts of   
 massive clusters \citep[e.g.][]{dolag06,porter08,mahajan12}, or superclusters\footnote{In this work a filament or a network of filaments bound by clusters of galaxies on opposite ends is referred to as a `supercluster'.} 
 \citep[e.g.][]{porter07,haines11,2018Smriti}, which often comprise two or more massive clusters and groups with large-scale filaments passing through them. In order to mimic such superclusters using the \e~simulations, in this work we identify systems where two massive clusters may be connected via filaments. 
 The \e~simulations however have a small box size (the largest box is 100 $\rm h^{-1}$ Mpc on each side), hence not many superclusters are expected to have formed in it \citep{2005Bagla,2009Bagla}. Therefore, in order to make the best use of the available data, we generated mock observations by identifying the positions of haloes with mass $\sim 10^{14} \ \rm M_{\odot}$. Nine such clusters were found, resulting in 36 ($\Comb{9}{2}$) potential slices comprising superclusters. Table~\ref{cluster_table} lists the properties of these haloes ($z=0.1$), which are analysed 
 further. To find the underlying structure of the cosmic web, we restricted ourselves to galaxies with $M_*/M_\odot \geq 10^9$ (within 30 pkpc aperture), which gives us a sample of 
 294,780 galaxies. This choice of stellar mass limit is motivated by earlier work which showed that for the RefL0100N1504 run, simulations are well matched to the observations for $M_{*}/M_\odot \geq 10^{9}$ \citep{EAGLE_paper,2016Tray}.

\begin{table}
    \caption{The physical parameters from the \e~simulation used to study the properties of galaxies. }
    \label{proptable}
    \begin{tabular}{lc}
        \hline
        Parameter & Description \\
        \hline
        $M_{*}$ & Total stellar mass within the aperture ($M_{\odot}$). \\
        $M_{gas}$ & Total gas mass within the aperture ($M_{\odot}$).\\
        $M_{tot}$ & Total subhalo Mass ($M_{\odot}$)\\
        x,y,z& Position of subhalo \\
        SFR & Star formation rate within the aperture.\\
        $u$ & u band without dust attenuation.\\
        $g$ & g band without dust attenuation.\\
        $r$ & r band without dust attenuation.\\
        \hline
    \end{tabular}
\end{table}

 \begin{table*}
     \caption{Clusters used for analysis. The columns are: (i) grpid: group id of the group detected by the friends-of-friends algorithm. (ii)  $M_{\rm tot}$: mass ($M_{200}$) of the group. 
     (iii) R: radius of the group halo. (iv) x, (v) y and (vi) z  are the respective coordinates of group halo, and, (vii) number of subhaloes identified by the SUBFIND algorithm. }
         \label{cluster_table}
    \begin{tabular}{|c|c|c|c|c|c|c|}
        \hline
        grpid & $M_{tot}/M_\odot$ & R (pkpc) & x (Mpc) & y (Mpc) & z (Mpc) & Number of subhaloes \\
        \hline 
        27000000000000&$3.4 \times 10^{14}$&1426.70&5.72&75.61&47.48&17101\\
        27000000000001&$1.7\times 10^{14}$&1135.12&18.12&79.88&53.51&9942\\
        27000000000002&$2.89\times 10^{14}$&1349.73&9.12&35.03&54.52&9152\\
        27000000000003&$2.9\times 10^{14}$&1356.48&52.43&5.00&20.12&7430\\
        27000000000004&$1.85\times 10^{14}$&1162.48&10.96&81.29&54.89&7365\\
        27000000000005&$1.38\times 10^{14}$&1054.56&77.85&76.97&42.28&6909\\
        27000000000006&$1.38\times 10^{14}$&1055.64&76.13&89.23&41.89&6191\\
        27000000000007&$1.11\times 10^{14}$&981.45&61.25&33.08&20.23&6183\\
        27000000000008&$1.00\times 10^{14}$&948.02&84.83&47.33&8.14&3190\\
        \hline
    \end{tabular}
      \end{table*}

\begin{figure}
    \includegraphics[scale=0.5]{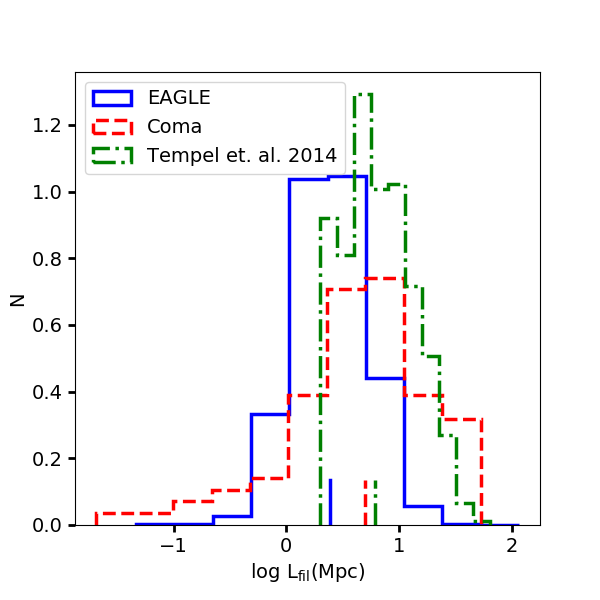}
    \caption{Comparison between the distribution of length of filaments in a study based on the SDSS data \citep{2014Tempel}, the Coma supercluster \citep{2018Smriti} and our sample. The vertical lines represent the median of the respective distributions. The filaments identified in both the observational datasets are found to be longer relative to the filaments from the \e~simulations in this work.}
    \label{plot:Lengthplot}
\end{figure}

\subsection{Mock Observations}
\label{Obsersec}

 Although simulations contain 3-dimensional (3D, henceforth) information, we created 2D projections of 3D slices in order to compare our results with the observations. Furthermore, 
 the 2D analysis allows us to artificially expand our dataset by rotating the central axis of the supercluster systems found in the simulation box as described below. 
 Motivated by observational studies of supercluster filaments crossing clusters of galaxies \citep[e.g.][]{dolag06,porter08,mahajan12,2018Smriti}, we begin by considering each cluster 
 pair system as a possible supercluster candidate. 
 
 In order to generate a slice of these data for making mock observations, the coordinate system is changed by translating the origin to the centre of the line joining the 
 two clusters in the system. The system is then rotated by keeping one axis fixed with the line joining the two clusters, and choosing the orientation of the remaining two axes randomly. 
 \citet{2018Laigle} have shown that a 60--200 Mpc thick slice can capture real 3D filaments even in their 2D projection. Therefore, 
 in order to characterise the environment in 2D projections, we choose a slice of co-moving width 60 Mpc with the normal perpendicular to the line joining the cluster pair as 
 shown in Fig. \ref{plot:setup}. This choice of width also compliments the width of the Coma supercluster region studied by \cite{mahajan12,2018Smriti}.
 In the following, we refer to this transformed frame as $\mathrm{O^\prime}$.

\subsection{Classification of Environment}
\label{sec:Class}

The cosmic web can be broadly segregated into different components based on the density of galaxies. Characterisation of these environments requires identifying the local topology and clustering
of the constituents. In this study, we follow an approach similar to that used in \citet[][hereafter SM18]{2018Smriti} to characterise environment into groups and clusters, filaments and voids, 
respectively. 

 In order to identify filaments in the 2D projected slices, we employ the Discrete Persistent Structures Extractor \citep[{\sc disperse};][]{2011Sousbie} algorithm. {\sc disperse} uses the
 Morse theory and persistence theory to map the persistence of critical points obtained by measuring density at discrete points to the underlying topology. It identifies features such as nodes, filaments, and saddle points using just one parameter. This parameter 
 called the persistence threshold crudely corresponds to the level of complexity to be retained while identifying structures in a distribution. We direct the reader to SM18 and \citet{2011Sousbie} for detailed application and working of the algorithm, and choice of parameters. 
 
  The persistence threshold above which all the critical points are retained for analysis is usually taken to be 3-5 $\sigma$, where $\sigma$ is the standard deviation. It suppresses the selection noise and picks relatively denser and longer filaments \citep{2011Sousbie}. In this work, we choose the significance threshold of $3\sigma$ which segregates the noise
  from the points used for analysis in this work for each mock observation.
 
 Filaments are obtained by applying {\sc disperse} on the distribution of galaxies in the $\mathrm{O^\prime}$ projection. In Fig.~\ref{proplot}(b) we show the filaments identified 
 for a representative slice. We identify filament galaxies as those which (i) do not belong to clusters or groups, and (ii) are $\leq 2$ Mpc from the spine of the filament. The choice of 
 filament radius is driven by the fact that $\sim 75\%$ of all galaxies found within $5$ Mpc of the spine of the filaments lie $\leq 2$ Mpc from the central axis (also see Sec.~\ref{s:colour}).  
 
 In order to identify the galaxies in each slice, we use \e's group catalogue to identify the friends-of-friends groups with total mass, 
 $M_{\rm halo} > 10^{12.5} \ \rm M_{\odot}$, and comprising at least four galaxies with $M_*/M_\odot > 10^9$ \citep{EAGLE_paper}. 
The galaxies which are not linked to groups, clusters or filaments were classified as void galaxies. In Fig.~\ref{finalexslice} we show the distribution of galaxies in different 
 environments in one of the slices.

\begin{figure}       
    \centering
    \includegraphics[width=\columnwidth]{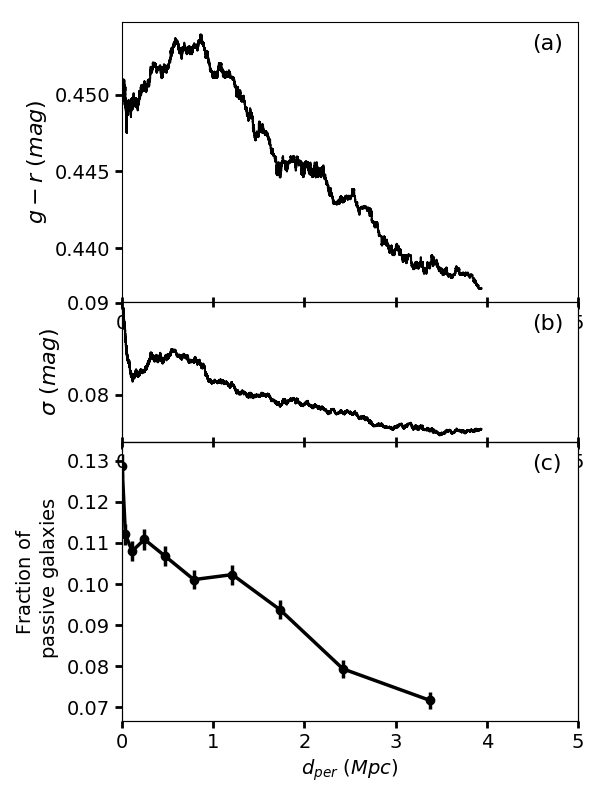}
    \caption{The (a) $g-r$ colour of
 galaxies, (b) median absolute deviation (MAD) of $g-r$ colour shown in (a), and (c) fraction of passive galaxies ($sSFR < 0.01$ Gyr) on filaments as a function of their distance from the spine of the filament. Galaxies become redder and the fraction of passive galaxies increases with decreasing distance from the spine of the filaments. }
\label{plot:dvscol}
\end{figure}

\section{Properties of galaxies on filaments}
\label{sec:Results}
 
 In order to study the properties of galaxies in different environments in the \e~simulations, we stack all the 36 slices obtained from mock observations. In the following we 
 analyse the trends in the properties of galaxies, in particular the galaxies on filaments with respect to their counterparts residing in denser and rarer regions of the space.

\subsection{The large-scale filaments}
\label{dpercolsec}

 In Fig. \ref{plot:Lengthplot} 
 we compare the distribution of length of filaments from our stacked sample generated from the \e~simulations, with the filaments in the Coma supercluster (SM18),
 as well as another distribution of filaments obtained from the SDSS dataset \citep[see fig.~11 of][]{2014Tempel}. While SM18 identified the filaments in 2D projections, just like this work, 
 the filaments of \citet{2014Tempel} were identified from the 3D distribution of galaxies.

 To compare these distributions, we use the Welch's t-test \citep{1947Welch}, which tests for the hypothesis that two distributions have different means. We find that the distribution of lengths
 of filaments from \e~simulation when compared with the filaments from SM18 and \citet{2014Tempel} gives p-values $< 0.01$ in both cases. 
 The filaments found in both the observational datasets are found to be relatively longer than the filaments found in the \e~simulations. However, this result can only be considered 
 qualitatively because of the different redshifts of the samples, and the fact that the procedure for finding the filaments is dependent on the algorithms with free parameters which 
 can influence the distribution of length of filaments \citep{libeskind18}.  
 
\subsubsection{Colours of galaxies}
\label{s:colour}

 Broadband optical colours are often a good representative of the age of the most dominant stellar population, and hence, the age of galaxies. Therefore, in order to study the impact of the filament 
 environment on the evolution of galaxies we analyse their colour as a function of increasing environmental density. In Fig. \ref{plot:dvscol} we show the running median of the $g-r$ colour of galaxies, median absolute deviation (MAD) of $g-r$ colour, and the fraction of passive galaxies \citep[sSFR $< 0.01$ Gyr$^{-1}$;][]{EAGLE_paper}, as a function of their perpendicular distance from the spine of filaments ($d_{\rm per}$). We find that the galaxies become redder and more passive closer to the spine of filaments within a radius of $2$ Mpc from the filament axis. Specifically, the passive fraction increases from 8\% at a distance of $\sim 2$ Mpc from the spine of the filaments to 11.5\% at the centre of the filaments. Welch's t-test statistic 
 shows that the colour of galaxies at $d_{\rm per} < 1$ Mpc is statistically different from their counterparts farther away from the spine of the filaments 
 (Table~\ref{stats}). The MAD quantifies the width of the distribution at different $d_{\rm per}$, essentially showing that the distributions are broader closer to the spine of the 
 filaments where most of the galaxies lie. 
This result is in tune with the findings from observations \citep{alpaslan16,2017Kuutma,2018Kraljic,2019Bonjean}, as well as other simulations \citep{2010Gay}. Such trends with $d_{\rm per}$ have helped in constraining the radius of filaments to $1-2$ Mpc \citep{2018Bonjean,2018Smriti,2019Tanimura}. 
 
\begin{figure}
    \includegraphics[width=1.0\columnwidth]{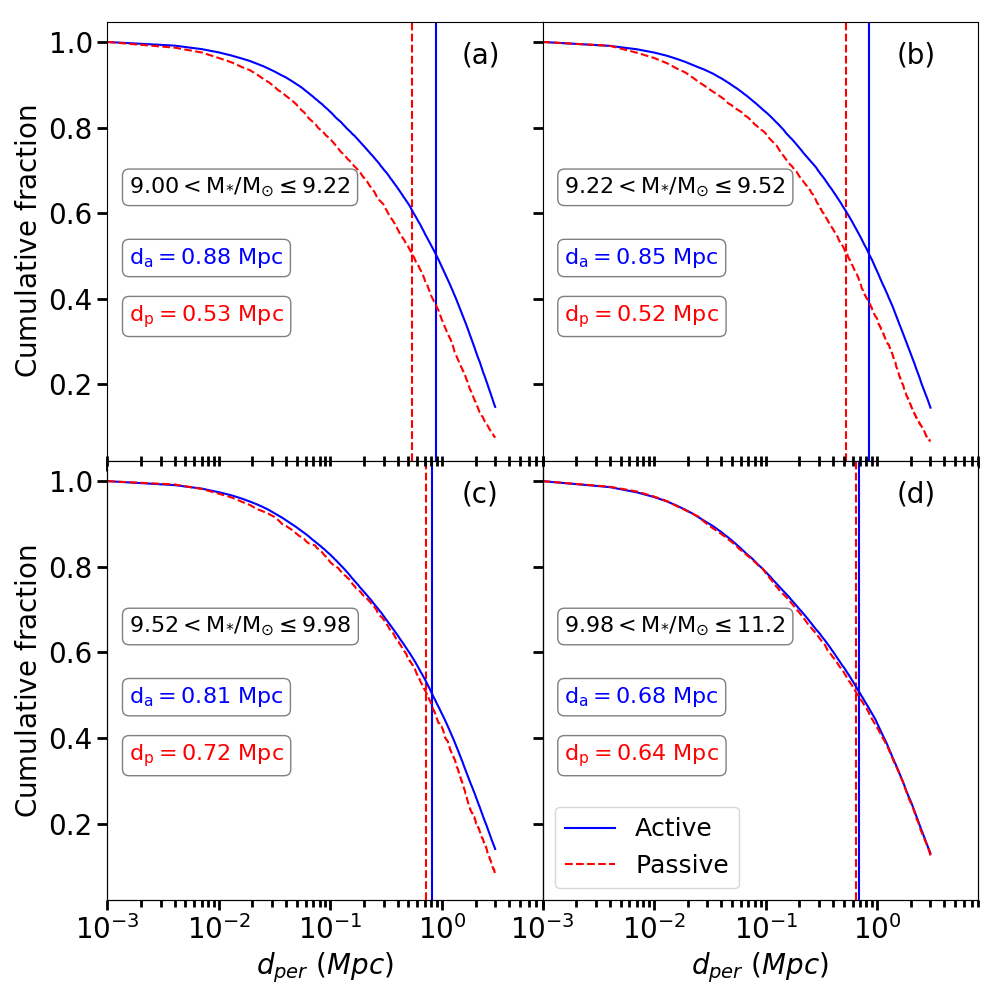}
    \caption{The cumulative distributions of active and passive galaxies are shown as a function of $d_{\rm per}$ for different mass bins. The {\it vertical dashed lines} correspond to median value for $d_{\rm per}$ for the passive ({\it dashed red line}) and the active ({\it solid blue line}), respectively. The four mass bins are chosen to have  the same number of galaxies in each bin. The median of the distribution for the active galaxies in the three low stellar mass bins (a)-(c) have value close to 1 Mpc, i.e. the radius of filaments comprising majority of the 
    galaxies on the cosmic web. The distributions for the active and passive galaxies in these three mass bins are also statistically distinct. On the other hand, the galaxies in
     the highest mass bin are statistically similar to each other. These distributions evidently show that 
    at fixed stellar mass, passive galaxies lie closer to the spine of the filament than their active counterparts.}
    \label{plot:fracapp}
\end{figure}

 The stellar mass of galaxies is well known to correlate with the environmental density. Hence in order to break this degeneracy between $M^*$ and environment, in Fig. \ref{plot:fracapp} we show the cumulative fraction for the active and passive galaxies in filaments separately in four mass bins: $10^{9}<M/M_{\odot} \leq 10^{9.22}$, (b) $10^{9.22}<M/M_{\odot} \leq 10^{9.52}$, (c) $10^{9.52}<M/M_{\odot} \leq 10^{9.98}$ and (d) $10^{9.98} < M/M_{\odot} \leq 10^{11.2}$. The mass bins are chosen to include the same number of galaxies in each bin. For the three low stellar mass bins ((a)-(c)), the median $d_{\rm per}$ for the active galaxies' distributions have values close to 1 Mpc, i.e. the radius of filaments comprising $\gtrsim 50\%$ of the galaxies on the cosmic web. We further tested for statistical differences among the distributions in each panel using the Kruskal-Wallis (KW) test \citep{1952Kruskal}. The KW statistic is a non-parametric rank-based test which assumes independent observations, continuous variability and same shape for the two distributions. 
 For the four panels in Fig.~\ref{plot:fracapp} the KW test yields a probability of 5.67e-82, 1.10e-52, 2.85e-06, and 0.31 in favour of the hypothesis that the two distributions had the same medians. Based on this analysis we also conclude that except for high mass galaxies, at fixed stellar mass, passive galaxies lie closer to the spine of the filament than their active counterparts. 
  
 \begin{table}
 \caption{The Welch test statistical probability for the likelihood that the galaxies in bin x are the same as bin y. The bins are chosen in ascending order of the distance 
 of galaxies from the spine of the filament such that, bin 1: $0 \leq d_{\rm per}/{\rm Mpc} \leq 1$, bin 2: $1 < d_{\rm per}/{\rm Mpc} \leq 2$ and bin 3: $2 < d_{\rm per}/{\rm Mpc} < 5$. }
 \begin{center}
 \begin{tabular}{c|c|c|c}     
 \hline
Parameter 	&    bin1-bin2   	& bin2-bin3   		& bin1-bin3  \\  \hline
$g-r$                 &  $1.497~\times~10^{-23}$    & $1.554~\times~10^{-27}$   & $3.059~\times~10^{-82}$ \\
sSFR 		&    0.310      	& $8.711~\times~10^{-25}$  		& $2.871~\times~10^{-44}$ \\
  $M_{gas}/M_{tot}$		&    0.058      	& $5.401~\times~10^{-14}$  		& $1.341~\times~10^{-31}$ \\
$M_*/M_{tot}$ 		&    $9.859\times~10^{-108}$ 	& 0.021      		& $5.094\times~10^{-158}$ \\
$Z_{NSF}$ 	&    $3.845~\times~10^{-127}$ 	& $2.321~\times~10^{-49}$  		& $\lll 0.0$ \\
$Z_{SF}$  	&    0.871     	& $1.216~\times~10^{-26}$  		& $5.194~\times~10^{-44}$ \\ 

\hline

\end{tabular}
\end{center}
\label{stats}
\end{table}

\subsubsection{Star formation in filament galaxies}

 The star formation rate of filament galaxies is enhanced relative to their counterparts in clusters as well as the voids \citep{fadda08,mahajan12}. SM18 showed that while the optical 
 and UV colours becomes bluer, the equivalent width of the H$\alpha$ line increases for the filament galaxies moving away from the spine of the filaments. 

 \begin{figure}
	\centering
	\includegraphics[width=\columnwidth]{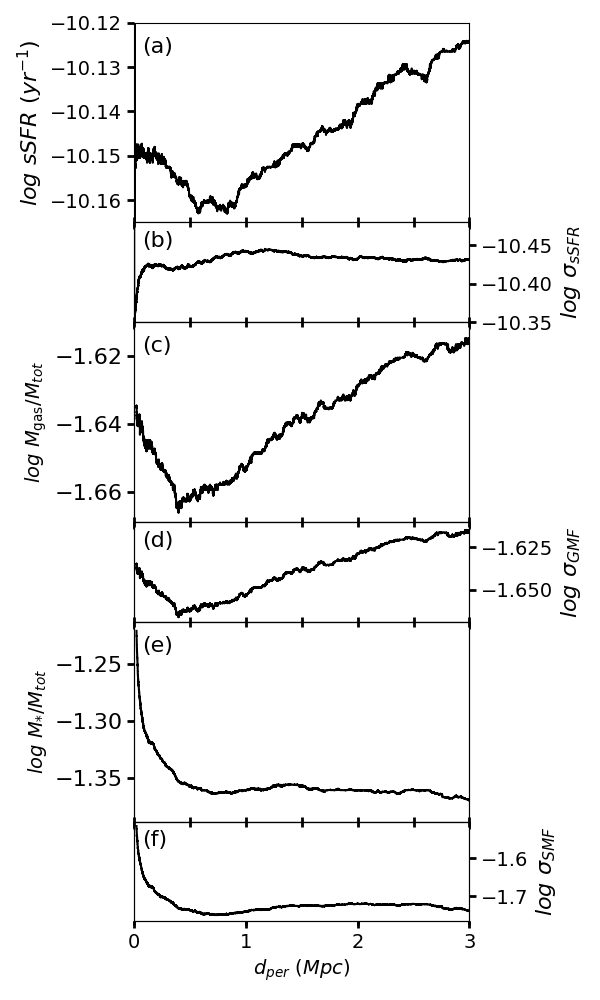}
	\caption{The (a) median sSFR, (c) gas mass fraction, and (e) stellar mass fraction as a function of $d_{\rm per}$. Plots (b), (d) and (f) show the MAD for sSFR, gas mass and stellar mass respectively. $M_{tot}$ is the sum of $M_{gas}$, $M_*$ and $M_{dm}$ inside the 30pkpc aperture of each galaxy. Outside a radius of 0.5 Mpc, the sSFR and the gas mass fraction decrease with decreasing distance from the spine of the filaments, while the stellar mass function remains constant. Within a radius of $\sim 0.5$ Mpc, the trend seems to reverse, and all three quantities exhibit enhancement with decreasing $d_{\rm per}$, indicating that the intrafilamentary medium condensing closer to the spine of the filament may be responsible for increasing the SFR of filament galaxies with $d_{\rm per} < 0.5$ Mpc. }
	\label{plot:dvsSFR}
\end{figure}

 In Fig. \ref{plot:dvsSFR}(a)-(b) we show the variation of the running median of the specific star formation rate (sSFR) as a function of the perpendicular distance ($d_{\rm per}$) to the spine of the filaments and the MAD for the same. On approaching the filaments, the sSFR decreases slowly until $\sim 1$ Mpc from the spine of the filaments, and thereafter increases again closer to the central axis. 
 Table~\ref{stats} confirms the observed trends by showing that the sSFR distribution for galaxies within $\lesssim 2$ Mpc is different from their counterparts further away from the 
 spine of the filaments.
 This result correlates well with the $g-r$ colour of galaxies becoming bluer in Fig. \ref{plot:dvscol}(a). Furthermore, these results support the findings of \citet{2019Liao}, who have suggested that the galaxies on filaments can accrete intrafilamentary gas to feed star formation, and the density of this gas should also increase with decreasing distance from the spine of the filaments.
 
 In order to test our hypothesis and test if the increase in star formation activity near the spine of the filaments is correlated with an increase in the availability of gas, 
 we study the distribution of mass in stars and gas in the filaments as a function of $d_{\rm per}$.
 In Fig. \ref{plot:dvsSFR}(c)--(f) we show the variation in the fraction of stellar mass and gas mass as a function of $d_{\rm per}$ for $\sim 96\%$ of the galaxies in our sample with non-zero gas mass fraction, and the respective MAD for the same. The denominator for the fractions, $M_{tot}$ is the total mass of the galaxy, which includes the gas ($M_{gas}$), stars ($M_{*}$) and the dark matter mass ($M_{dm}$) inside a 30 pkpc aperture. 
 
 The median stellar mass fraction (SMF) shows a mild increase only within $\lesssim 0.5$ Mpc from the central axis of the filaments. This trend is reciprocated by the gas mass fraction (GMF) as well. However, at $d_{\rm per}>0.5$ Mpc, the median GMF rises smoothly, but the median SMF remains constant. The observed trends are also confirmed statistically in
 Table~\ref{stats}. Furthermore,   
 we note that all the trends seen in Fig. \ref{plot:dvsSFR} are primarily driven by low mass ($10^{9} \leq M/M_{\odot} \leq 10^{10}$) which dominate our sample. Galaxies with stellar 
 masses in the range $10^{10}-10^{10.5}$ also show trends similar to their less massive counterparts with minor deviations. But the most massive galaxies show an 
 almost constant distribution for sSFR and the gas mass fraction, but an incline in the stellar mass fraction distribution with decreasing $d_{\rm per}$ at small 
 distances ($d_{\rm per} \lesssim 0.5$ Mpc) from the spine of the filament. 

 These trends suggest that the intrafilamentary gas condenses into the galaxies on filaments and fuel star formation, supporting the results of \citet{2019Liao}.
 However, these small changes (2\%--5\%) in the GMF and the sSFR are not enough to turn a passively evolving galaxy to an active one as seen here in Fig.~\ref{plot:dvscol}, and in 
 other studies based on observations \citep{2019Bonjean}. Also because of the small magnitude of these changes it is not surprising that such trends have yet not been observed in real data. The increase in the passive fraction of galaxies is likely caused by increased gravitational interactions amongst filament galaxies, likely due to an increased number density of galaxies closer to the spine of the filament. As a result of these interactions, these galaxies not only exhaust their reservoir of gas, but will also show momentary increase in their star formation activity, therefore contributing to the trend observed for the sSFR. Similar conclusions have also been drawn by \cite{2019Bonjean}, who modelled the passive fraction of 
 filament galaxies to find a positive gradient in their SFR which they attributed to mergers.

\begin{figure}
    \includegraphics[width=\columnwidth]{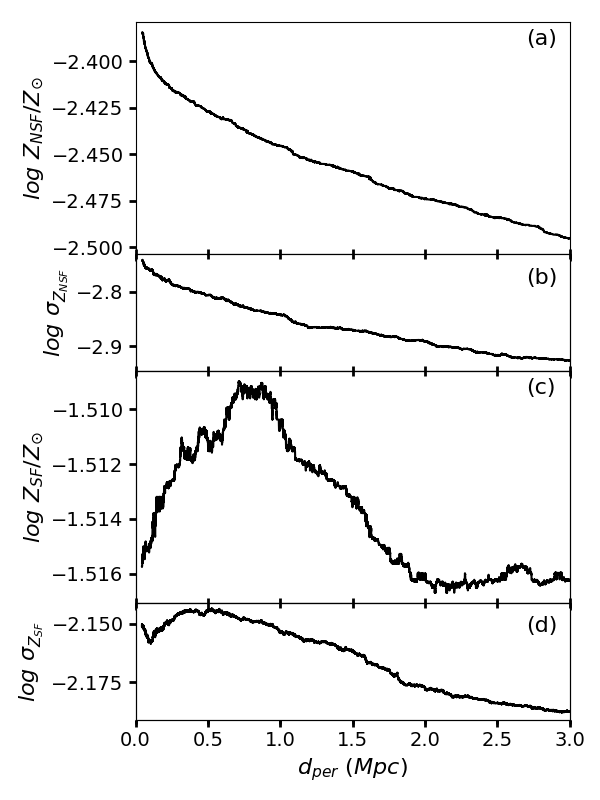}
    \caption{The median metallicity of (a) the star forming gas ($Z_{SF}$) and, (c) the non star-forming gas ($Z_{NSF}$) as a function of $d_{\rm per}$. Plots (b) and (d) show MAD for $Z_{SF}$ and $Z_{NSF}$ respectively. 
    The $Z_{SF}$ within $d_{\rm per} \lesssim 1$ Mpc is elevated relative to gas further away. The  $Z_{NSF}$, on the other hand increases smoothly with decreasing $d_{\rm per}$, indicating accretion of non-star-forming gas on the filament galaxies.}
    \label{plot:dpervsZ}
\end{figure}

\subsubsection{Metallicity}
\label{subsec:Metallicity}

 Metallicity, the fraction of mass in elements heavier than Helium is dependent on the complex interplay of gas accretion on galaxies and outflows from different feedback processes. For our analysis of metallicity, we use the smoothed metallicity provided for 
 each subhalo by \citet{2009Wiersma}. The gas metallicity is divided into the star-forming ($Z_{SF}$) and non-star-forming ($Z_{NSF}$) metallicity, respectively. The star-forming gas is cold ($\sim 10^{4} K$) enough to form stars, with the SFR dependent upon the metallicity \citep{2008Schaye}. On the other hand, the non-star forming gas comprises hot and 
 under-dense gas which can not form stars. We refer the reader to \citet{2004Schaye} for a detailed discussion on how star-formation is implemented in the \e~simulation.

 Our prior results indicate that galaxies are likely to accrete gas as they move towards the spine of the filament. Therefore, in order to test this hypothesis using 
 $Z$ we show the running median of $Z_{SF}$ as a function of their perpendicular distance from the spine of the filaments, $d_{\rm per}$ in Fig.~\ref{plot:dpervsZ}. Although with significant scatter, the median 
 $Z_{SF}$ within $\lesssim 2$ Mpc of the spine of the filaments is clearly elevated relative to the galaxies further away. On the other hand, $Z_{NSF}$ is found to increase smoothly with decreasing distance from the filament axis\footnote{Just like Fig. \ref{plot:dvsSFR}, these trends in metallicity are also driven by the low-mass 
 galaxies ($10^{9} \leq M/M_{\odot} \leq 10^{10.5}$). The more massive galaxies show noisy, almost constant distribution of both $Z_{SF}$ and $Z_{NSF}$ except an 
 apparent decline within $d_{\rm per} \lesssim 1$.}  (Fig. \ref{plot:dpervsZ}(c)). 
 This result, together with Fig. \ref{plot:dvsSFR}(c) indicates that the non-star forming gas is accreted 
 onto the galaxies. This result also supports the findings of \citet{2019Liao} who used hydrodynamical simulations to show that at high redshifts ($z = 4.0$ and $2.5$) around 
 30\% of the gas accreted on the galaxy halos on filaments is the intrafilamentary medium.

\section{Discussion}
\label{sec:Discussion}
We have used data products from the EAGLE simulations to show that the intermediate density environment prevalent on the filaments influence the properties of galaxies on them. In the following, we discuss our results in the context of the filament galaxies studied in the literature, and also analyse the properties of filament galaxies relative to their counterparts in other environments. 
 
\subsection{Filaments in the literature}

\cite{2017Kleiner} used the 6 degrees Field Galaxy Survey \citep{2009Jones} and the Parkes all-sky survey \citep{2000Stavenley} to study the HI fraction of galaxies on nearby filaments identified by {\sc disperse}. They found that massive galaxies ($M_{*}/M_{\odot} >= 10^{11}$) closer ($< 0.7$ Mpc) to the spine of the filaments have higher HI-to-stellar mass ratio 
 (HI fraction), where as the HI fraction among lower mass galaxies show no difference with respect to the control sample of galaxies $> 5$ Mpc away from the spine of the filaments.
 \cite{2017Kleiner} attributed the higher HI fraction in massive filament galaxies to cold gas accretion from the intrafilamentary medium. 

 On the other hand, \cite{2018Crone} used data from the ALFALFA survey \citep{2005Giovanelli} to study the HI reservoir in galaxies. Complimentary to the results presented here, 
 \cite{2018Crone} observed that at fixed stellar mass and colour, galaxies show a deficiency in HI on approaching the filaments. Furthermore, they also observed that at fixed local 
 density, the HI fraction of lower mass galaxies ($10^{8.5}<M_{*}/M_{\odot}<10^{10.5}$) decreases on approaching the filaments. 
 
\begin{figure}
	\includegraphics[width=\columnwidth]{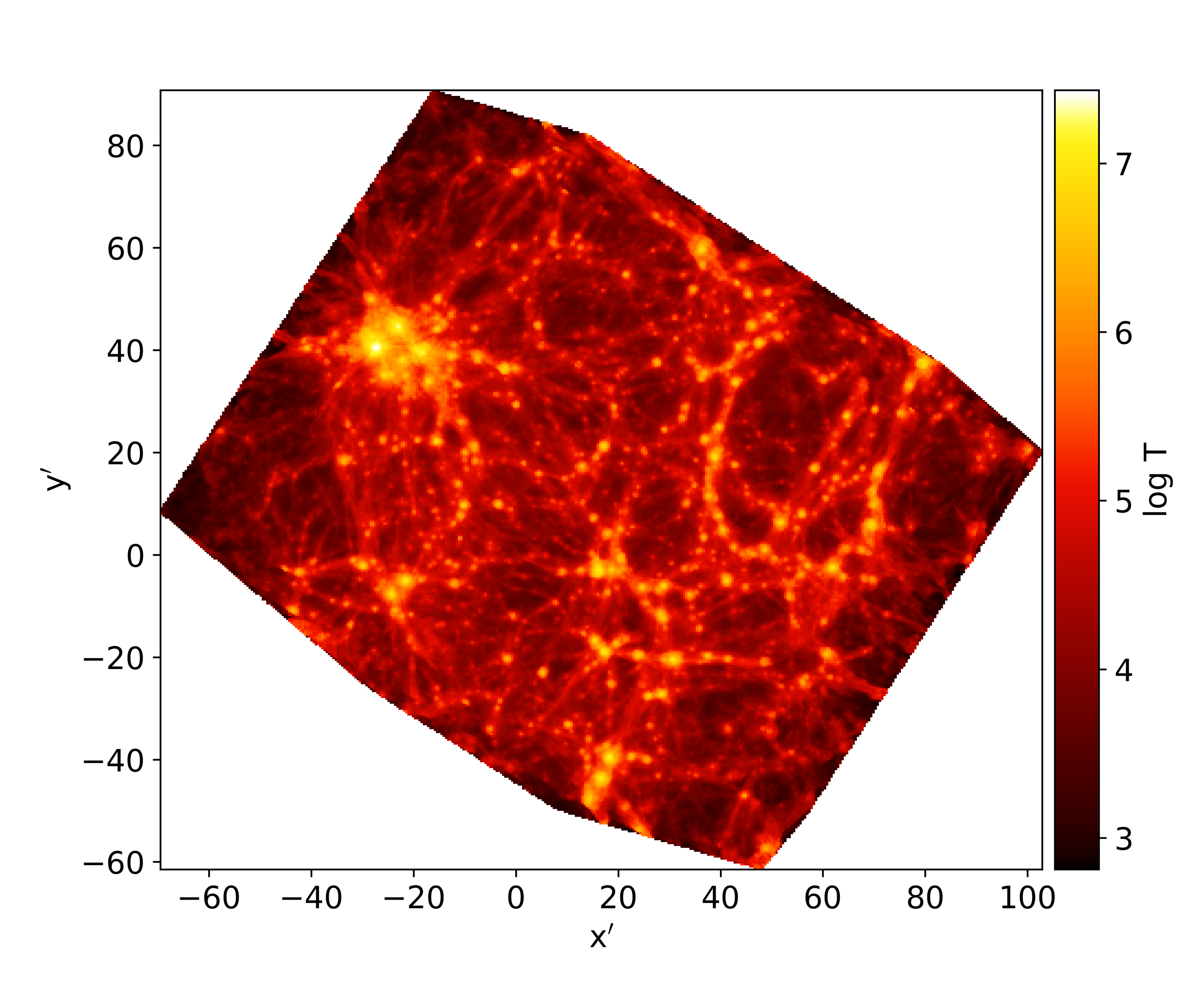}
	\caption{The projected temperature for the sample slice plotted in Fig. \ref{finalexslice}. This plot makes use of the particle dataset at $z=0.1$. The colours represent the mean of $log T$. The filaments clearly show concentration of gas at $10^{4}<T(K)<10^{5}$. This gas can be accreted by the filament galaxies.} 
	\label{plot:temp}
\end{figure}

 Even though the lower mass limit of our sample is 0.5 dex greater than the sample used by \cite{2018Crone}, we can make use of the particle dataset to study the trends observed in the \e~simulation. In Fig.~\ref{plot:dvsSFR} we show that the gas mass fraction decreases with $d_{\rm per}$ until $~ 0.5$ Mpc from the spine of the filament, but at $d_{\rm per} \lesssim 0.5$ Mpc, the trend seems to reverse. It is worth noting that the stellar mass of galaxies also seems to increase within the same radius around the spine of filaments. These results, together with enhanced sSFR within $0.5$ Mpc from the spine from the filament (Fig.~\ref{plot:dvsSFR} (a)), indicate that the enhanced gas condensing in galaxies closer to the spine of the filaments may lead to an increase in the rate of formation of stars. This hypothesis is in tune with the results of \citet{2017Kleiner} as well as Fig. \ref{plot:temp}, where we have used the particle dataset from the \e~simulation to show the projected temperature map for the gas in the example slice shown in Fig. \ref{finalexslice}. It is evident that the filaments in the \e~simulation comprise gas at $10^{4}<T(K)<10^{5}$, which can fuel star formation in galaxies, especially closer to the spine of the filaments \citep{2019Bonjean}. 
 
\cite{2015Darvish} studied 28 galaxies on filaments against 30 galaxies in voids from the COSMOS survey \citep[$z\sim0.53$;][]{2007Scoville}.
They found that the star-forming galaxies on filaments have higher metallicity than their counterparts in the voids. \cite{2015Darvish} attributed this difference to the accretion of pre-enriched filamentary gas on to the filament galaxies. In a complimentary study of filaments at $z=1.5$ from the Horizon hydrodynamical simulation, \cite{2010Gay} also found that the 
 gas-phase metallicity of galaxies increases with decreasing distance from the spine of the filaments. 
Our analysis shows similar trends in Fig. \ref{plot:dpervsZ}. However, we also note that the \e~simulations' reference model used here (RefL0100N1504) is known to over-predict the metallicity relative to runs with smaller box lengths and observations \citep[fig. 1;][]{2017DeRossi}. But since this over-prediction is consistent throughout the box, it should not influence the trends observed in the metallicity as a function of environment.    

\cite{2017Kuutma} used the SDSS data to study the statistical variation in the properties of galaxies on approaching the spine of the filaments. 
 \cite{2017Kuutma} found that the galaxies become more elliptical and redder in the $g-i$ colour with decreasing distance from the spine of the filaments. Using a similar dataset, SM18 studied the Coma supercluster to find a similar increase in the $g-r$ colour index of galaxies with decreasing $d_{\rm per}$. A similar trend is reflected in the analysis presented here in Fig.~\ref{plot:dvscol}. We find that the median $g-r$ colour of galaxies becomes redder with decreasing distance from the spine of filaments. 

 \citet{alpaslan16} analysed 1799 star-forming spiral galaxies ($z\leq0.09$) within $3.79~h^{-1}$ Mpc of the centre of filaments using data from the Galaxy and Mass Assembly 
 (GAMA) survey. In agreement with the results presented in this work as well as other literature \citep[e.g.][]{2018Kraljic,laigle18}, they found that galaxies closer to the spine 
 of the filaments are more massive, and at fixed \smass, galaxies are quenched closer to the spine of the filaments. Based on our own results and those presented in the 
 literature we therefore suggest that along with stellar mass and large-scale environment, anisotropic tides along filaments impact the assembly history of galaxy halos, and 
 hence the properties of galaxies.

\subsection{Filaments vs groups and voids}
\label{sec:compare}

In this section, we compare the properties of galaxies on filaments to their counterparts in groups and clusters, and the voids. 

\begin{figure}
    \centering
    \includegraphics[width=0.9\columnwidth]{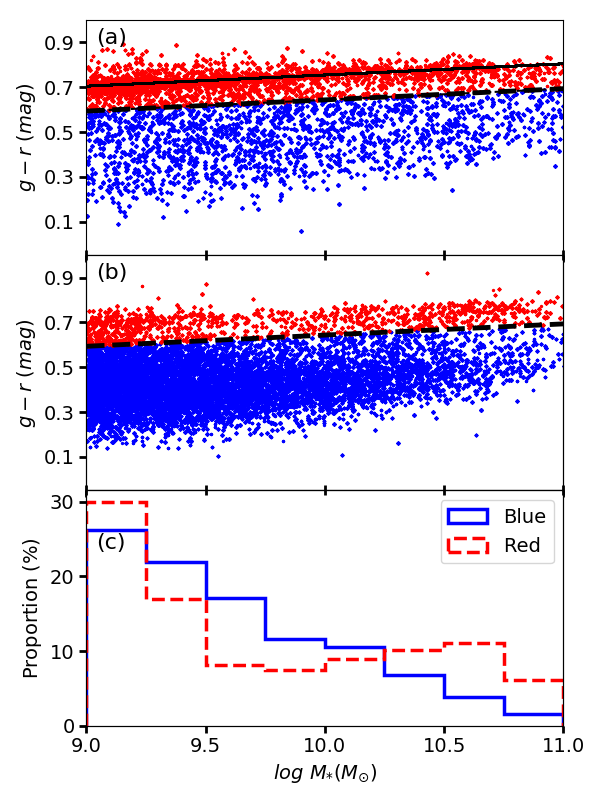}
    \caption{The (a) $g-r$ colour of  galaxies in groups and clusters, and (b) filaments as a function of their stellar mass. The black line shows the red sequence fitted to the galaxies in clusters. In this work all galaxies bluer than $2\sigma$ from the red sequence are considered as blue, and the complimentary population is considered red. (c) This figure shows the kernel density estimates of stellar mass for the red and the blue populations on filaments. The Welch's t-test gives a p-value of $2.25e-175$, indicating that the two populations are statistically distinct. It is also evident that although majority of blue galaxies have masses $< 10^{10} M_{\odot}$, a significant fraction of red galaxies follow suit. }
    \label{plot:BlueRed}
\end{figure}

In Fig.~\ref{plot:BlueRed} we show the distribution of galaxies in the colour-$M_*$ plane for the groups and filaments, respectively. In order to compare the two colour distributions, we fit the red sequence to the distribution of cluster galaxies. In this work, all galaxies with $g-r$ colour more than $2\sigma$ bluer than the red sequence are classified as blue galaxies, where $\sigma$ is the standard deviation (0.055 mag) obtained by iteratively fitting the red sequence. It is evident from 
Fig.~\ref{plot:BlueRed} (b) that there are fewer red galaxies on filaments, relative to the clusters. Specifically, while $\sim 54\%$ of the cluster galaxies are red, only $13\%$ of the filament galaxies follow suit. 

In Fig.~\ref{plot:BlueRed}(c) we plot the kernel density estimates (KDE) of stellar mass for galaxies in both the colour selected populations of filament galaxies. KDE are closely related to histograms but are continuous and smooth. KDE is a non-parametric methods for calculating probability distribution of a variable using a non-negative function called kernel. We used Gaussian kernel and scott estimator \citep{2015Scott} for bandwidth used in the kernel. A large fraction of the blue galaxies (77\%) are found to have stellar mass, $M_{*}/M_{\odot}\le 10^{10}$. On the other hand, while most of the red galaxies on filaments are massive, around $63\%$ of them have $M_{*}/M_{\odot}\le 10^{10}$. 
The Welch's t-test applied to the two distribution yields a p-value of $2.25e-175$, inferring that the two populations are statistically distinct. This result suggests that some environmental mechanism(s) may be responsible for quenching star formation in lower mass galaxies on filaments. 

\begin{figure}
    \centering
    {\includegraphics[width=0.8\columnwidth]{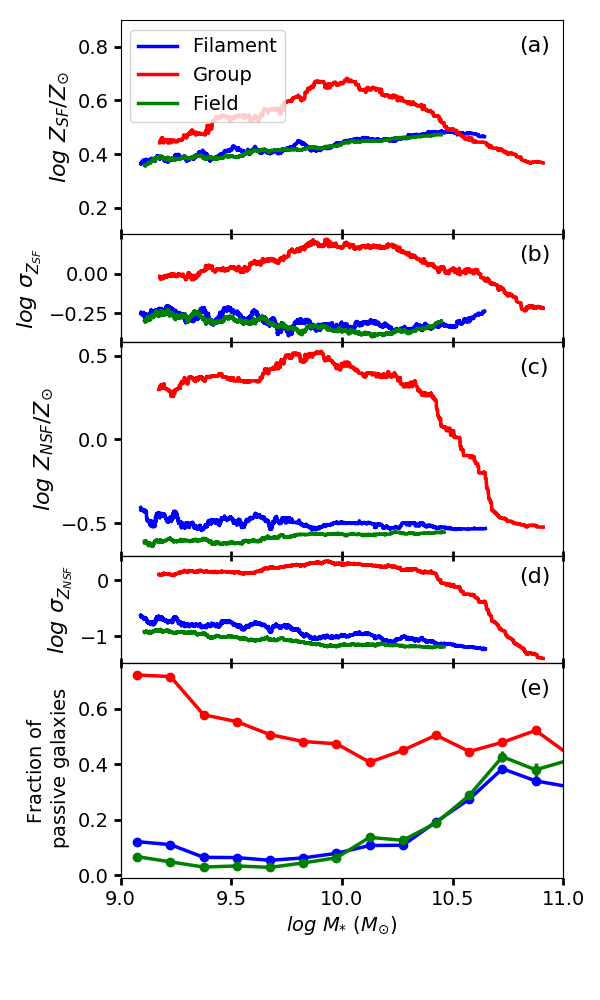}}
    \caption{The (a) median metallicity of star forming gas ($Z_{SF}$) and (c) non star-forming gas ($Z_{NSF}$)as a function of stellar mass ($M_{*}$) of galaxies in groups and clusters ({\it red}), filaments ({\it blue}) and voids ({\it green}), respectively. MAD for $Z_{SF}$ and 
    $Z_{NSF}$ is shown in (b) and (d) respectively. (e) The fraction of passive galaxies as a function of $M_{*}$ in the three environments. In groups the fraction of passive galaxies increases with decreasing $M_{*}$ for galaxies with $M_{*} < 10^{10} M_{\odot}$. While an opposite trend is observed for the filaments and voids. }
    \label{plot:EnvMstar}
\end{figure}

 The metallicity of a galaxy is related to its stellar mass $M_{*}$, such that more massive galaxies also have higher gas as well as stellar metallicities \citep{1968McClure,1979Lequeux,2002Garnett,2004Temonti,2006Lee,2010Mannucci,2019Cresci}. In Fig.~\ref{plot:EnvMstar}(a and c) we show the median metallicity of the star-forming gas ($Z_{\rm SF}$), and the non-star forming (NSF) gas ($Z_{\rm NSF}$) as a function of $M_*$ in different environments. In agreement with the previous findings \citep{2017DeRossi}, the NSF gas is found to be less metal rich relative to the star-forming gas. While the galaxies on filaments and voids show almost constant $Z_{\rm SF}$ and $Z_{\rm NSF}$, their counterparts in groups show 
 marginally different trends on either side of $M_{*}/M_{\odot} \sim 10^{10}$, such that the Welch t-test probability is 0.045 in favour of the hypothesis that the distribution of $Z_{\rm SF}$ 
 for the lower mass galaxies ($M_{*}/M_{\odot} \lesssim 10^{10}$) is statistically similar to their massive counterparts. The respective probability for $Z_{\rm NSF}$ for group 
 galaxies is found to be $\ll 0$.  
 Such trends of lower $Z$ in the most massive galaxies have also been observed in the Coma cluster \citep{pogg01}, and the Coma supercluster \citep{tiwari20}.  

Fig.~\ref{plot:EnvMstar} can be compared to fig.~1 of \cite{2017DeRossi}, who have also made use of the \e~simulation to compare the mass-metallicity relation obtained in simulations with various observational datasets and found them to agree within uncertainties. Our analysis indicates that the slope of the mass-metallicity relation is primarily determined by the galaxies in dense environments.

In Fig.~\ref{plot:EnvMstar}(c) we show the fraction of passive galaxies ($sSFR < 0.01 \ \rm Gyr^{-1}$) as a function of stellar mass in different environments. It is evident that in the groups' environment, the fraction of low-mass ($M_{*}/M_{\odot}<10^{10}$) passive galaxies increases significantly with decreasing $M_{*}$. On the other hand, the galaxies in filaments and voids show identical trends in the opposite direction, i.e. the fraction of passive galaxies increases with $M_{*}$, particularly for galaxies with $M_{*}/M_{\odot}>10^{10}$. This result is in broad agreement with the literature \citep[e.g.][]{haines06} where the dwarf galaxies are found to be passive only if they are a satellite to a massive galaxy. SM18 have also shown that the $NUV-r$ colour distribution of galaxies changes from voids to clusters, mainly because of a higher fraction of massive galaxies turning redder in dense environments.

\begin{figure}
	\includegraphics[width=\columnwidth]{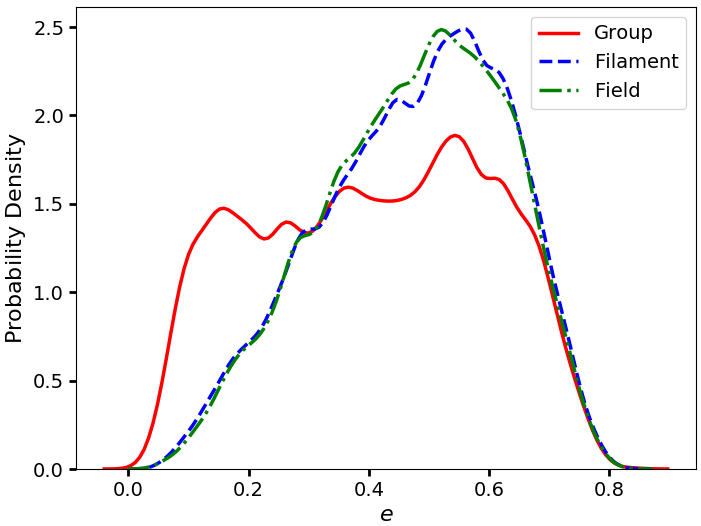}
	\caption{The distribution of ellipticity (e) of galaxies in the red population in different environments. Galaxies in clusters show a wider range of ellipticities relative to their counterparts in filaments and voids. The Welch's t-test for the distribution of $e$ of galaxies in all but one pair of populations is $\sim 0$. A p-value of 0.116 for the galaxies in voids and filaments indicate that the morphology of these two populations is statistically identical.}
	\label{plot:ellip}
\end{figure}

The morphology of galaxies can provide important clues to their evolution. In this work, we use the morphology of resolved galaxies having more than 300 stellar particles as per the procedure described by \cite{2019Thob}. The shape parameters are derived using the reduced iterative inertia tensor in ellipsoidal apertures. Ellipticity is defined as:
\begin{equation} 
e=1-\frac{c}{a}
\end{equation} 
where, $c$ and $a$ are the length of semi-minor and semi-major axis, respectively. In Fig.~\ref{plot:ellip} we plot the distribution of $e$ for galaxies in different environments. The Welch's t-test for all but one pairs of environments yield p-value of $\sim 0$. The distribution of $e$ for the filament and void galaxies has the Welch's t-test probability of 0.116 in favour of the hypothesis that the two distributions are identical, suggesting that the ellipticity of the void and filament galaxies are statistically indistinguishable. This result is in agreement with the observations \citep{2015Alpaslan} where the GAMA survey data was used to show that the galaxies in filaments and voids have similar distributions of ellipticity, such that the maximum of the distribution lies in the range $0.2-0.6$. 
On the other hand, the groups and clusters are dominated by low $e$, i.e. more elliptical galaxies. 

\section{Conclusion}
\label{sec:Conclusion}

We employ data from the \e~simulations to explore properties of galaxies in different environments, in particular the large-scale filaments. We summarize the major findings of our analysis below:
\begin{itemize}
    \item Galaxies become redder and more passive closer to the central axis of the filaments. Furthermore, this trend is likely observed due to an increase in the relative fraction of passively-evolving galaxies having $M_*/M_\odot \lesssim 10^{10}$ (Figs.~\ref{plot:fracapp}, \ref{plot:BlueRed}(c)) closer to the centre of the filaments. In agreement with the results from various observations, this trend is observed within a radius of 2 Mpc around the spine of the filament, thus providing an upper limit for the filaments' radius.
    
    \item Both, the gas and stellar mass fractions of galaxies rise closer to the spine of the filaments within $d_{per}<0.5$ Mpc. But at $d_{per}>0.5$ Mpc the SMF remains constant while the GMF rises smoothly with increasing distance from the spine of the filaments. Together with the findings from the literature, these results lead us to conclude that the observed trends are a consequence of increased gravitational interactions between filament galaxies closer to the spine of the filament. This enhancement in the interaction rate is caused by an increased number density of galaxies closer to the centre of the filaments.
    
    \item The morphology of galaxies on filaments is similar to their counterparts in voids, but statistically distinct from the ones in clusters. However, fewer red galaxies are found on the filaments relative to clusters. 
\end{itemize}

 To conclude, our results indicate that many properties of galaxies on filaments are different from their counterparts in clusters and voids. The colour, star formation, metallicity and gas fraction of filament galaxies change as a function of their distance from the central axis of the filament. This could be an impact of the IFM as well as the number density of galaxies, both of which increase with decreasing $d_{\rm per}$. High resolution spectroscopic and HI data for galaxies covering all environments and a range of stellar masses are required to confirm our results. We hope the upcoming facilities such as the Vera Rubin Observatory (VRO) would be crucial for the same. At shorter wavelengths, the Advanced Telescope for High ENergy Astrophysics (ATHENA) x-ray observatory is likely to unravel the properties of the warm hot intergalactic medium (WHIM) in the 0.5--12 kev range.  

\section{Data availability}

The data underlying this article were derived from sources in the public domain: \url{http://icc.dur.ac.uk/Eagle/database.php} 

\section{Acknowledgement}
We are grateful to the reviewer for their suggestions which have greatly improved this manuscript.
The authors would like to acknowledge the high-performance computing facility at IISER Mohali. We acknowledge
the Virgo Consortium for making their simulation data available. The \e~simulations were performed using the DiRAC-2 facility at Durham, managed by the ICC, and the 
PRACE facility Curie based in France at TGCC, CEA, Bruy\`{e}res-le-Ch\^{a}tel. Singh thanks Pooja  Munjal, IISER Mohali, for her help in making Fig.~\ref{plot:setup}.
 Mahajan is funded by the INSPIRE Faculty award (DST/INSPIRE/04/2015/002311), Department of Science and Technology (DST), Government of India. 




\bibliographystyle{mnras}
\bibliography{Refs} 

\label{lastpage}

\end{document}